\documentclass[%
 aip,
 jmp,%
 amsmath,amssymb,
 reprint,%
]{revtex4-1}
\usepackage{amsmath}
\usepackage{amsfonts}
\usepackage{graphicx}
\usepackage{hyperref}
\usepackage{mathtools}
\usepackage{verbatim}
\usepackage{subfigure}
\usepackage{hyperref}
\usepackage{latexsym}
\usepackage{dcolumn}
\usepackage{epsf}
\usepackage{float}
\usepackage{ulem}
\DeclarePairedDelimiter\abs{\lvert}{\rvert}

\usepackage{graphicx}
\usepackage{dcolumn}
\usepackage{bm}
\usepackage{color,soul}

\begin{document}

\title{Amplification of explosive width in complex networks}
\author{Pitambar Khanra} \affiliation{Department of Mathematics,
National Institute of Technology, Durgapur 713209, India}
\author{Prosenjit Kundu} \affiliation{Department of Mathematics,
National Institute of Technology, Durgapur 713209, India} \author{Pinaki
Pal} \affiliation{Department of Mathematics, National Institute of
Technology, Durgapur 713209, India}
\author{Peng Ji}
\affiliation{The Institute of Science and Technology for Brain-inspired
Intelligence, Fudan University, Shanghai, China} \author{Chittaranjan
Hens}
\affiliation{Physics \& Applied Mathematics Unit, Indian Statistical
Institute,  Kolkata 700108, India}

\date{\today}
 
\begin{abstract} 
We present an adaptive coupling strategy to induce hysteresis/explosive synchronization (ES) in complex networks of phase oscillators Sakaguchi-Kuramoto model). The coupling strategy ensures explosive synchronization with significant explosive width enhancement.  Results show  the robustness of the strategy and the strategy can diminish (by inducing enhanced hysteresis loop) the contrarian impact of phase frustration in the network, irrespective of network structure or frequency distributions. %
 Additionally, we design a set of frequency for the oscillators which eventually ensure complete  in-phase synchronization behavior among these oscillators (with enhanced explosive width) in the case of adaptive-coupling scheme.
 Based on a mean-field analysis, we develop a semi-analytical formalism, which can accurately predict the backward transition of synchronization order parameter.


\end{abstract}

\maketitle

\begin{quotation}
The phenomenon of synchronization has fascinated researchers for many years due to its appearance in variety of natural as well as manmade systems. However, the study of synchronization in adaptively coupled complex networks has got less attention.  
Here,  we have presented an adaptive coupling scheme and explored the impact of that in the enhancement of explosive width in Sakaguchi-Kuramoto model on complex networks. Numerical investigation shows that the proposed scheme ensures explosive synchronization (ES) with significant explosive width enhancement for diverse frequency distributions  with different network realizations. More importantly, it is found that the proposed coupling scheme can inhibit the contrarian impact of the phase frustration in the network. We also have established a semi analytical treatment for investigating the system using the Ott-Antonsen ansatz. The results obtained from the semi analytical approach are found to match closely with the numerical simulation results. 
 The analysis shows that the adaptive coupling function is robust and can enhance the hysteresis width  in different complex networks having variety of natural frequency distributions and for a broad range of frustration parameter.
\end{quotation}

\section{Introduction} Adaptively coupled oscillators have  been found to exhibit a wide range of complex behaviors \cite{Aoki2009, Ha2016,
Kasatkin2018,Brede2019}.  For instance, link weights co-evolving with underlying dynamics can enhance the synchronizability in heterogeneous
networks \cite{Chen2009} and  this mechanism can induce  spontaneous synchronization in e.g., certain group of neurons of a brain by 
adjusting the adaptive Hebbian learning process \cite{Chakravartula2017,Niyogi2009,Skardal2014, Li2010}. 
Recently, it has been reported that the adaptation of the coupling configuration  by the local  order parameter may result in explosive synchronization, a phenomenon characterized by discontinuous (first-order) phase transitions between incoherent and coherent states accompanied by hysteresis in networks of coupled oscillators~\cite{Khanra2018, Kachhvah2017,Danziger2019, Danziger2016,Jesus2011,Pinto2015,Boccaletti2016,Leyva2013}.\\
\begin{figure}
\includegraphics[height=!,width=0.48\textwidth]{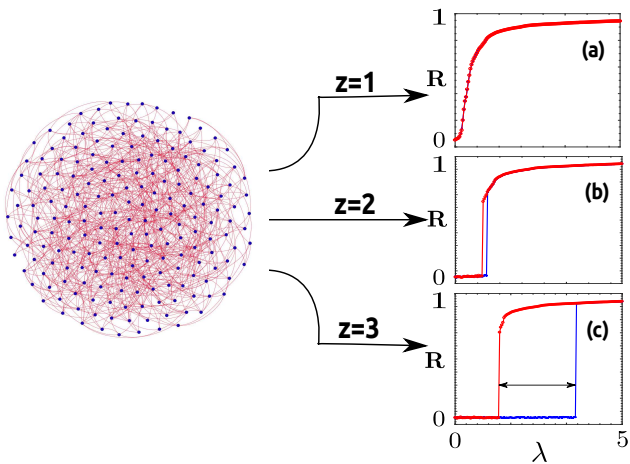}
\caption{{\bf Expansion of hyeteresis width in small network
}Visualization of the expansion of hysteresis width for ER network of
network size $N = 300$, $\left\langle k \right\rangle  =12$ 
  with increasing $z$. Right panel synchronization diagram is
  constructed using the model Eqn.\ (\ref{eqn1}) with phase frustration value $\alpha=0.5$. Double headed arrow in the last figure of right panel indicates explosive width of the synchronization transition.} \label{Figure1}
  \end{figure} 
Of particular interest, the explosive synchronization \cite{Zhang2015, Filatrella2007} has been found to emerge among adaptively-coupled oscillators irrespective of the structure of the  networks or intrinsic frequency distributions.
We raise here the question  what happens  if we use a modified strategy, i.e  incorporating the high power of magnitude of the adapting parameter?

Here, we have used the modified adaptive-couping scheme proposed by Filatrella et al.~\cite{Filatrella2007} on Sakaguchi-Kuramoto model in complex network environment, and results show that the proposed scheme can amplify or increase the hysteresis width (difference between the backward transition and forward transition points of synchronization) for diverse frequency distributions with different network realizations. 
For demonstration, we start with the classical network-coupled Sakaguchi-Kuramoto phase oscillators of size $N$
\cite{Kuramoto1984,Acebron2015,Dorfler2012,Rodrigues2016}. 
Dynamics of each oscillator in the network is governed by the equation
\begin{eqnarray}\label{eqn1} 
\frac{d\theta_i}{dt} &=& \omega_i + \lambda
l_i \sum_{j=1}^{N} A_{ij}\sin(\theta_j - \theta_i-\alpha),
\end{eqnarray} 
where $\omega_i$ is  the intrinsic  natural frequency of
the $i^{th}$ oscillator, $A_{ij}$ is an $ ij^{th}$ element of the
adjacency matrix $A = (A_{ij})_{N\times N}$.
Here $\alpha$ accounts for a frustrating term   within the range of
$0\leq \alpha < \frac{\pi}{2}$
\cite{Sakaguchi1986,Omel'chenko2012,Skardal2015,Skardal2015_PhysD,Lohe2015} and $\lambda$ is the coupling strength. $l_i$ is a time-dependent coupling term and  contributes  adiabatically to the coupling function.    
 
As a general choice, we consider $l_i$ as a function of the  Kuramoto order parameter $R$ which
quantifies the degree of synchronization of the entire population. 
 The global order parameter is
defined as 
\begin{eqnarray}\label{Global}
 Re^{i(\psi-\alpha)} =
\frac{1}{N}\sum_{j=1}^{N}e^{i(\theta_j-\alpha)}, 
\end{eqnarray} 
where
${0}\leq{R}\leq{1}$ quantifies the magnitude of coherence and $\psi$
denotes the average phase. 
Prior research have shown that the adaptive
function $l_i=R$ can result in explosive synchronization 
\cite{Zhang2015,Khanra2018}. 
However, this strategy generates a constant
 hysteresis or explosive width. We seek here a suitable choice of $l_i$  which can enhance the hysteresis loop of  phase-coupled networks  and  choose the adaptive function as $l_i=R^{z-1}$, where $z$ ($\geq1$) is a  positive real number~\citep{Filatrella2007} . 
 \par To illustrate the
effectiveness of our choice,  we consider a heterogeneous
Erd\H{o}s-R\'{e}nyi (ER) network with $N= 3 \times 10^2$  nodes. 
We increase (decrease) the coupling 
strength $\lambda$ adiabatically with an increment (decrement)
$\delta{\lambda}=0.01$ and compute the stationary value of $R$ for each
$\lambda$ during the forward (backward) transition from the incoherent
to the phase synchronized state. In this simulation we take the natural
frequencies from a Lorentzian distribution and set the frustration parameter
($\alpha$) equal to 0.5. In Fig.\ \ref{Figure1} (a), the order
parameter undergoes a continuous transition for the choice of $z=1$
where the network  is a purely diffusive   i.e. the coupling function is
not controlled by any adaptive coupling. However, setting the  adaptive
parameter to higher power ($z=2$), one can generate discontinuous
(explosive)  transition in the synchronization order parameter (see Fig.\
\ref{Figure1} (b)). Interestingly, the explosive width is further enhanced
if we increase the value of $z$ from $2$ to $3$ (Fig.\ \ref{Figure1}
(c)). The double headed arrow in Fig.\ \ref{Figure1}(c) indicates the explosive width for the explosive synchronization transition which is the difference between the backward transition and forward transition points of synchronization.
In this paper,  we have explored the  impact of $z$ on transition to synchronization in different complex networks of Sakaguchi-Kuramoto oscillators for a  broad range of   frequency distributions, namely Lorentzian or uniform distributions. We have  also considered the degree-frequency correlated environment and a special form of degree-frequency correlation which eventually gives perfect synchronization. Our results show that the adaptive coupling function is robust and can induce the enhanced explosive width for a broad range of frustration parameter over diverse network settings. In this work, we validate our  detailed numerical results by solving (semi-analytically) order parameter equation on the basis of Ott-Antonsen ansatz~\cite{Ott2008}. The forward transition curve  is perfectly fitted with our semi-analytical findings.
  
\section{Analytical Approach and Numerical verification}
We start with the \textit{annealed network} approximation proposed in~\cite{Dorogovtsev2008, Coutinho2013,Kundu2017,Ichinomiya2004,Peron2012}, which gives critical behavior of phase transitions (including synchronization transition) in complex networks. For the case of a sparse uncorrelated complex network with a degree distribution $P(k)$  in the thermodynamic limit ($N\rightarrow\infty$) we may write,
  \begin{eqnarray}\label{annealed_eqn} 
  \frac{d\theta_i}{dt} &=& \omega_i
  +\frac{\lambda R^{z-1} k_i}{N\left\langle k \right\rangle }
  \sum_{j=1}^{N} k_{j}\sin(\theta_j - \theta_i-\alpha), 
  \end{eqnarray}
where $k_i$ is the degree of $i^{th}$ node, and $\langle k \rangle$ is the average degree across nodes. To obtain  the explicit set of equations for the time evolution of the complex order parameter $C=Re^{i(\psi-\alpha)}$, we use here the Ott-Antonsen ansatz~\cite{Ott2008} where the density of the oscillators  at time $t$ with phase $\theta$ for a given degree $k$ and frequency $\omega$ be given by the function $f(k,\omega;\theta,t)$, and we normalize it as 
  \begin{eqnarray}
   \int_{0}^{2\pi}
  f(k,\omega;\theta,t)d\theta =
  h(k,\omega)=P(k)g(\omega).\label{density} 
  \end{eqnarray} 
  To maintain the conservation of the oscillators \cite{Kuramoto1984,Ichinomiya2004}
  in a network, the density function $f$ satisfies the continuity
  equation 
  \begin{eqnarray}\label{continuity} 
  \frac{\partial}{\partial
  t} f(k,\omega;\theta,t) + \frac{\partial}{\partial \theta}[v
  f(k,\omega;\theta,t)) = 0, 
  \end{eqnarray} 
  where $v$ is the velocity 
  field on the circle that drives the dynamics of $f$, which comes from
  the right hand side of the Eqn.\ (\ref{annealed_eqn}) as follows
  \begin{eqnarray} 
  v(k, \omega; \theta, t)= \frac{d\theta}{dt}= \omega +
  \frac{\lambda k}{2i}R^{z-1}\left[ Ce^{-i\theta}-C^*e^{i\theta}\right].  
  \end{eqnarray}\label{v_eqn} 
  
  Expanding $f(k, \omega; \theta, t)$ in a
  Fourier series in $\theta$, we have\\ 
  \begin{eqnarray}
  f=\frac{h(k,\omega)}{2\pi}\left\lbrace 1 + \left[
  \sum_{n=1}^{\infty}f_n(k,\omega,t)e^{i n \theta} + c.c.\right]
  \right\rbrace, 
  \end{eqnarray} 
  where c.c. stands for complex conjugate. 
  We now consider the Ott-Antonsen ansatz 
  \begin{eqnarray}\label{ansatz}
  f_n=\beta^n(k,\omega,t) 
  \end{eqnarray} 
  to obtain an equation for the   function $\beta(k,\omega,t)$ and find the following equation
  \begin{eqnarray}\label{beta_diff} 
  \frac{\partial \beta}{\partial t} +
  i \omega \beta + \frac{\lambda k}{2}R^{z-1}(C\beta^2 - C^*)=0,
  \end{eqnarray} 
  where $\abs{\beta(k,\omega,t)}\leq 1$ to avoid
  divergence of the series. 
  The above equation is not  yet in a closed form. 
  So to make the above equation in a closed form, we  get the
  complex global order parameter as 
  \begin{eqnarray}\label{c(t)_eqn}
  C(t)=\frac{e^{-i\alpha}}{\left\langle k
  \right\rangle}\int_{k_{min}}^\infty
  \int_{-\infty}^{\infty}kP(k)g(\omega)\beta^* d\omega dk.
  \end{eqnarray}

Now to study a steady state solution we average 
$C(t)=Re^{i\psi-i\alpha+i\Omega t}$ for the constant order parameter
$R$, a phase $\psi$, and a group angular velocity $\Omega$. 
By suitably varying the reference frame, $\omega\mapsto \omega-\Omega$ and putting
$\psi=0$, without loss of generality, we have $C=Re^{-i\alpha}$. Therefore,
$C^*=Re^{i\alpha}$ and for stationary points of Eqn.\ (\ref{beta_diff})
we find the solution of $\dot{\beta}=0$. 
\begin{figure}
\includegraphics[height=!,width=0.48\textwidth]{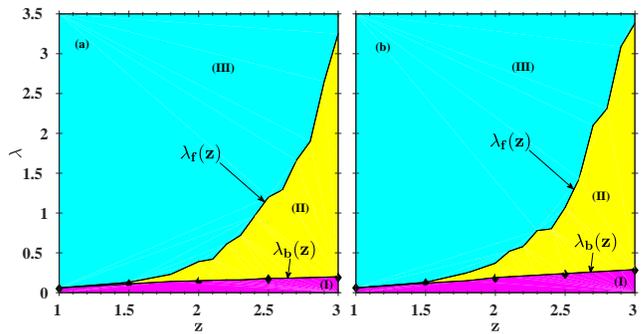} 
\caption{{\bf Phase diagram on $z-\lambda$ plane showing amplification of hysteresis width.}
Magenta (I), yellow (II) and cyan (III) islands respectively represent
asynchronous ($R\sim 0$), hysteresis and synchronous ($R\sim 1$)
regions.  These regions are separated by solid black line indicating the
critical coupling strength for the transition to synchrony during the
forward $(\lambda_f)$ and backward $(\lambda_b)$ continuation. Both the
figures are constructed with a scale free (SF) network of size $N=2000$, $\left\langle
k \right\rangle  =12$ and $\gamma=2.8$. Panel $(a)$ indicates zero phase
frustration i.e., $\alpha=0$ and panel $(b)$ indicates $\alpha=0.5$. The solid black boundary separating regions (II) and (III)  and  solid line separating the regions (I) and (II) has been obtained from  numerical simulation.  The black diamonds on this line are obtained from semi-analytical approach (Eqn.\ \ref{real_r}- \ref{im_r}) showing very close match.}
\label{Figure2} 
\end{figure} 

We obtain the solution as follows,
\begin{eqnarray} 
\beta(k,\omega)=\begin{cases}
	-i x e^{i\alpha}+e^{i\alpha}\sqrt{1-x^2}, & \text{$\abs x \leq 1$}.\\
	-i x e^{i \alpha}\left[1-\sqrt{1-\left(\frac{1}{x}\right)^2}\right], & \text{$\abs x > 1$}.
	\end{cases}
\end{eqnarray} 
where $x=\frac{\omega-\Omega}{\lambda k R^z}$ and this
solution  satisfies the requirement $\abs {\beta} \leq 1$ for the 
convergence of the geometrical series (Eqn.\ (\ref{ansatz})).  The first
solution corresponds to the synchronous state and the second solution is
due to desynchronous state. In this case, the order parameter can be rewritten
as \begin{eqnarray}
R = \frac{1}{\langle k\rangle} \bigg [\int \int_{k_{min}}^{\infty} k
P(k) g(\omega) \beta^*(k,\omega) H\left(1-\abs x\right)dk d\omega
\nonumber \\ + \int \int_{k_{min}}^{\infty} k P(k)
g(\omega)\beta^*(k,\omega) H\left(\abs x-1\right) dk d\omega \bigg ],~~~
\label{r_d_l}
\end{eqnarray}

\noindent where $H$ is Heaviside functions. The first part of right-hand
side of Eqn.\ (\ref{r_d_l}) encompasses the contribution of locked
oscillators  and the second part accounts for the contribution of drift
oscillators  to the order parameter $R$. Splitting the contributions of
real  and imaginary parts, we can eventually obtain the coupled self
consistent equations of $R$ and $\Omega$ (see in the appendix).

To test the  impact of $z$ on transition to synchronization, we use a heterogeneous scale-free network ($N = 2000$, $\left\langle k \right\rangle  =12$ and $\gamma=2.8$) generated from Barab\'{a}si-Albert model,  the frequencies are drawn from a Lorentzian distribution with mean $0$ and standard deviation $1$. The bi-stable state (hysteresis regime) is shown by yellow color (regime:II)  in the Fig. \ref{Figure2} (a)-(b) for the frustration  parameters $\alpha=0.0$ and $\alpha=0.5$ respectively. For lower $z$, the hysteresis width is found to be negligibly small and if we increase this adaptive  parameter to higher values,  the width is significantly enhanced  irrespective of the frustration parameter. Here the magenta island (regime: I) shows the de-synchronized regime and cyan island (regime: III) is  the locked part of coupled phase  oscillators. It is observed from the figure that as $z$ is increased, the region of the hysteresis is amplified significantly. Interesting to note here that the higher adaptive coupling parameter ($z > 1$) not only induces explosive synchronization in the network but also enhances the explosive width even for high phase frustration (e.g. $\alpha = 0.5$ in Fig.\ \ref{Figure2} (b)).  Typically in this kind of network, higher phase frustration ($\alpha > 0.3$) destroys the hysterisis behaviour in a degree-frequency correlated network when $z = 1$~\cite{Kundu2017}.  However, for higher value of $z$, it can overcome the situation and explosive synchronization (ES) re-emerges.


The boundary of the yellow and cyan regions which signifies the forward critical point ($\lambda_f (z)$) 
of the hysteresis transition is computed numerically from the model Eqn.\ \ref{eqn1}. 
The backward transition points derived from the self-consistent coupled equations of 
$R$ and $\Omega$ (see the appendix) are shown with black diamonds in Fig.\ \ref{Figure2} which show close match with numerical simulation.

\begin{figure}
\includegraphics[height=!,width=0.48\textwidth]{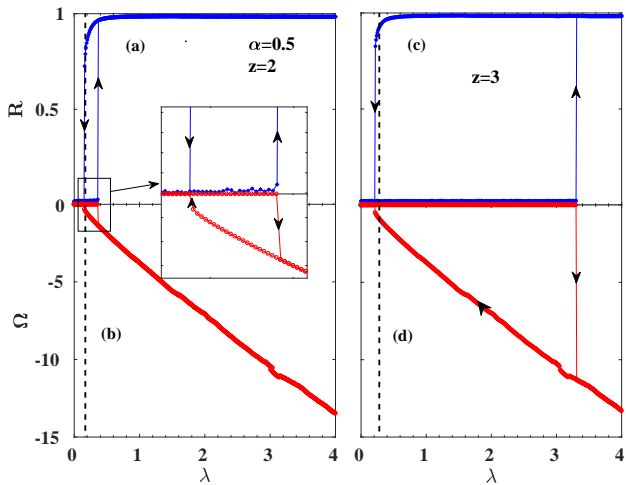}
 \caption{{\bf Order parameter $R$ and group
angular velocity $\Omega$ as a coupling strength $\lambda$ for two
values of $z$.} Blue curve indicates the synchronization diagram and red
curve indicates the group angular velocity calculated from the model
Eqn.\ (\ref{eqn1}) for SF network of network size $N = 2000$,
$\left\langle k \right\rangle  =12$ and $\gamma=2.8$. Black dotted verticle line indicates the backward transition point of $R$ and $\Omega$ calculated from the semi-analytic Eqn.\ (\ref{real_r}) and (\ref{im_r}). All the data are
simulated with phase frustration $\alpha=0.5$.} \label{Figure3}
\end{figure}

We have also numerically calculated the order parameter and the  global frequency $\Omega$ for $z=2$ and $z=3$ (see Fig.~\ref{Figure3}). The order parameter ($R$) and global frequency ($\Omega$) are shown in the Fig.~\ref{Figure3}(a,c) with blue lines and Fig.~\ref{Figure3}(b,d) with red lines respectively as a function of $\lambda$. The backward transition points determined from the self-consistent equations are shown with dashed vertical lines in the figure which shows a very close match with the backwards points obtained through numerical simulation of the whole system.

For $z=2$, the hysteresis width (see inset of Fig.\ \ref{Figure3} (a,b)) occurs due to the presence of adaptive coupling. For $z=3$ the width is
significantly enhanced compared to $z=2$. Note that, the values of $\Omega$ and $R$  are negligibly small before the forward critical coupling. The reason is as follows: adaptive function $R^{(z-1)}$, acting on each node of  the network, decreases  the effective  coupling ($\Lambda_{eff}=\lambda\times R^{(z-1)}$ where $0<R<1$ and $z>1$) substantially. Therefore, higher strength is required to establish locked phase.  Numerical values of  backward $R$ and $\Omega$
closely match with the ones obtained from the semi-analytical expression of $R$ and $\Omega$ derived from the Eqn.\ (\ref{r_d_l}). 
\begin{figure}
\includegraphics[height=!,width=0.48\textwidth]{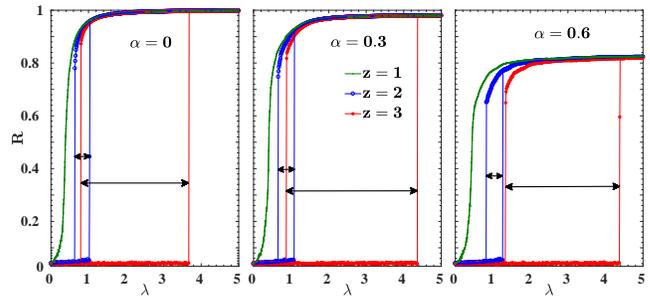}
\caption{{\bf Realization of the expansion of explosive width with z for three different values of phase frustration $\alpha$.} Both the figures are constructed using Watts-Strogatz small world network of size $N = 500$, $\left\langle k \right\rangle  =4$, and $\beta=0.5$. Double headed arrow indicates the explosive width of the hysterisis transition.} \label{Figure4}
  \end{figure} 
We therefore establish the fact, that  a tuning parameter associated with adaptive term  in power can enhance the width of  hysteresis  irrespective of the frustration value or network structure. For better realization we also simulate our model Eq.\ {\ref{eqn1}} in Fig. (\ref{Figure4}) with Watts-Strogatz small world network size $N=5\times10^3$, $\left\langle k \right\rangle  =4$, and $\beta=0.5$ for three different values of phase frustration $\alpha=0, 0.3, \text{and}~0.6$. In every panel of Fig. (\ref{Figure4}) we have simulate the model with increasing z.   We perform the numerical analysis for diverse set of frequency distributions, and we  show that 
the enhancement properties  appear for all the cases which only depend on the $z$. The changes in forward and backward points  depend on the
choices of frequency distributions. Next, we  validate our proposition in frustrated networks for a certain choice of frequency distribution directly correlated with its own degree in which a global order parameter achieves {\it perfect synchronization}\cite{Kundu2018} ($R= 1$). This is important, as $z$ mainly contributes in the enhancement of hysteresis width irrespective of $\alpha$. Although $\alpha$ frustrates the system by not allowing the order parameter to reach perfect synchronization. We seek a selection of frequency distribution which can avoid the impact of frustration $\alpha$ by reaching perfect synchronization ($R=1$) as well as it will create a broader explosive width in presence of higher values of $z$.

\section{Amplification of hysteresis widths in degree-frequency environment}

We have already shown that our adaptive strategy  works
efficiently when the frequencies are drawn from Lorentzian
distribution. A degree-frequency environment naturally induces ES
\cite{Jesus2011,Coutinho2013}, however a frustrated network
($\alpha>0.3$ \cite{Kundu2017}) cannot reveal ES even in the presence of
degree-frequency correlated dynamics. Here we seek to understand the
impact of such coupling configuration in  frustrated dynamics.

\begin{figure}
\includegraphics[height=!,width=0.48\textwidth]{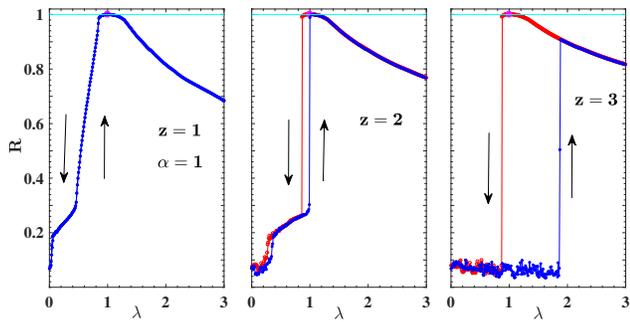} 
\caption{{\bf Amplification of explosive
width in case of degree-frequency correlation.} Order parameter $R$ as a
function of coupling strength $\lambda$ for a SF network of network size
$N = 2000$, $\left\langle k \right\rangle  =14$, and $\gamma=2.8$ for
different values of $z$. Blue line indicates the forward transition and
red line indicates backward transition. Magenta dot on the cyan color
line indicates our targeted point at $(1,1)$} \label{simu:rand_N1000}
\end{figure}

We consider a SF network of size $N=2000$ with $\gamma=2.8$ and average
degree $\langle k \rangle=14$, where natural frequency of each
oscillator scales with its own degree as $\omega_i=ak_i$ where $a$ is a
proportionality constant.  We set the phase-frustration parameter
$\alpha=1$, and we take  $a=\sin\alpha$. In this case order
parameter $R$ does not exhibit the first-order transition for
non-adaptive environment $(z=1)$ (Fig.\ \ref{simu:rand_N1000}, shown in
left column with blue line). Interestingly one additional characteristic
appears in  the behavior of the  order parameter, i.e., $R$ approaching $1$ at
$\lambda=1$. This is due to the emergence of perfect synchronization
($R=1$)  \cite{Kundu2018,Brede2016} for choosing a specific kind of
frequency distribution.

 As we know, in perfect synchronization state,
$\abs{\theta_j-\theta_i}= 0$ and $R= 1$. Using Taylor's series approximation,
 the Eqn.\ (\ref{eqn1}) can be written as
 \begin{eqnarray}
 \frac{d\theta_i}{dt}=\tilde{\omega_i}-\lambda \cos(\alpha)  \sum
 L_{ij}\theta_j, 
 \end{eqnarray}
 where $\tilde{\omega_i}=\omega_i-\lambda
 \sin(\alpha)k_i$ and $L$ is the Laplacian matrix. In vector form, one
 can write $ \boldsymbol{\dot \theta}=\tilde{\boldsymbol\omega}-\lambda
 \cos(\alpha) L {\boldsymbol\theta}$. Assume that, in global synchronization,
 all the oscillators follow a common frequency $\Omega$ and if we consider a
 $\Omega$ rotating frame, the phases will be freezed in this frame by
 setting themselves into steady states ($\frac{d
 {\boldsymbol\theta}}{dt}=0$). This further implies
 ${{\boldsymbol\theta}^*=\frac{L^\dagger{\bf\tilde{\boldsymbol\omega}}}{
 \lambda \cos(\alpha)}}$, where $L^\dagger$ is the pseudo-inverse
 operator of the Laplacian matrix $L$. Now if we choose $\omega_i=\sin(\alpha) k_i$ which sets ${\bf\tilde{\boldsymbol\omega}} =0$   
  at $\lambda=1$, and that eventually  gives ${\boldsymbol\theta}^*=0$.
 This signifies that the  frequency ensures perfect synchronization of
 oscillators ($R=1$) at $\lambda=1$ (Magenta dot in each panel of the
 Fig.\ \ref{simu:rand_N1000} on the cyan line).
Although the choice of the higher value of $z$ ($=2$) in $R^{z-1}$ 
yields the explosive width (Fig.\ \ref{simu:rand_N1000} middle panel). 
The   hysteresis width  is significantly increased for  higher
values of $z$ ($=3$) (shown in the right panel).
Therefore, we have shown that our adaptive strategy can create ES and
enhance the ES width in these types of networks irrespective of the choice
of the frequencies. 
The analytical prediction is confirmed in Fig.\
\ref{simu:rand_N1000} assuring perfect synchronization at $\lambda=1$
which also nullifies the contrarian  effect of frustration parameter
$\alpha$. The forward line misses the perfect synchronization point
(blue line) where as backward line coincides with $R=1$ at $\lambda=1$ for each
cases of $z$. 

\section{Conclusions}
We have investigated the phenomenon of explosive synchronization in complex networks of phase oscillators both numerically and analytically, based on an adaptive coupling strategy. Numerical simulations for different networks both in frustrated and non frustrated environment show that, as a parameter ($z$) in the coupling function is tuned appropriately, networks undergo abrupt transition to (explosive) synchronization and width of the associated hysteresis loop is greatly enhanced. It is observed that the proposed strategy is quite general and is applicable for any type of networks  and frequency distributions for the amplification of hysteresis width. We have confirmed that the strategy can successfully induce the emergence and enhancement of explosive synchronization in diverse frustrated environments. Based on the Ott-Antonsen ansatz, we have  derived analytical expressions for the global order parameter and global frequency. The results obtained from the semi-analytical approach based on Ott-Antonsen ansatz match with the backward transition points obtained from the numerical simulation of the entire complex networks. 

\section{Acknowledgements} PK acknowledges support from  DST, India under the DST-INSPIRE scheme (Code: IF140880). PJ acknowledges support from National Key R\&D Program of China (2018YFB0904500), NSF of Shanghai, Eastern Scholar and by NSFC 269 (11701096). CH is supported by
INSPIRE-Faculty grant (Code: IFA17-PH193).

\section{APPENDIX: Coupled equation of $R$ and $\Omega$} The
contribution of locked oscillators to the order parameter is
\begin{widetext} \begin{eqnarray} R_{l}=\frac{(\cos \alpha -i\sin
\alpha)}{\langle k\rangle}\int_{k_{min}}^{\infty} \int k P(k) g(\omega)
\left[\sqrt{1-\left(\frac{\omega-\Omega}{\lambda k R^z}\right)^2} +
i{\frac{\omega-\Omega}{\lambda k R^z}}  \right]dk d\omega
H\left(1-\abs*{\frac{\omega-\Omega}{\lambda k R^z}}\right)
\label{r_lock} \end{eqnarray} Now the contribution of drift oscillators
to the order parameter is given by \begin{eqnarray} R_{d}=\frac{(\sin
\alpha +i\cos \alpha)}{\langle k\rangle}\int_{k_{min}}^{\infty} \int k
P(k) g(\omega)\frac{\omega-\Omega}{\lambda k R^z}
\left[1-\sqrt{1-\left(\frac{\lambda k R^z}{\omega-\Omega}\right)^2} 
\right]dk d\omega H\left(\abs*{\frac{\omega-\Omega}{\lambda k
R^z}}-1\right) \label{r_drift} \end{eqnarray}

 \end{widetext}

 Hence we get $R=R_l +R_d$, where $R_l$ and $R_d$ are given by
 Eqn.(\ref{r_lock}) and Eqn.(\ref{r_drift}) respectively.

 Now comparing the real and imaginary parts we get \begin{widetext}
 \begin{eqnarray} R \langle k\rangle= cos \alpha \int_{k_{min}}^{\infty}
 \int k P(k) g(\omega) \sqrt{1-\left(\frac{\omega-\Omega}{\lambda k R^z}
 \right)^2} H\left(1-\abs*{\frac{\omega-\Omega}{\lambda k R^z}}\right)
 dk d\omega + \frac{sin \alpha}{\lambda R^z}(\langle \omega \rangle -
 \Omega) \nonumber \\-\sin \alpha \int_{k_{min}}^{\infty} \int  k P(k)
 g(\omega)\frac{\omega-\Omega}{\lambda k R^z}
 \sqrt{1-\left(\frac{\lambda k R^z}{\omega-\Omega}\right)^2}
 H\left(\abs*{\frac{\omega-\Omega}{\lambda k R^z}}-1\right) dk d\omega
 \label{real_r} \end{eqnarray} and \begin{eqnarray} \langle \omega
 \rangle -\Omega = \lambda R^z \tan \alpha \int_{k_{min}}^{\infty} \int
 k P(k) g(\omega) \sqrt{1-\left(\frac{\omega-\Omega}{\lambda k
 R^z}\right)^2} H\left(1-\abs*{\frac{\omega-\Omega}{\lambda k
 R^z}}\right) dk d\omega \nonumber \\ +\int_{k_{min}}^{\infty} \int P(k)
 g(\omega)(\omega-\Omega) \sqrt{1-\left(\frac{\lambda k
 R^z}{\omega-\Omega}\right)^2} H
 \left(\abs*{\frac{\omega-\Omega}{\lambda k R^z}}-1\right) dk d\omega
 \label{im_r} \end{eqnarray} By solving the above two equations we can
 get the set of values for order parameter $R$ corresponding to coupling
 strength $\lambda$. \end{widetext}


\begin{thebibliography}{28} \bibitem{Aoki2009} T. Aoki, and T. Aoyagi,
\textit{Phys. Rev. Lett.} {\bf 102}, 034101 (2009). \bibitem{Ha2016} S-Y
Ha, S. E. Noh, and J. Park, \textit{SIAM} {\bf 15}, 162-194 (2016).

\bibitem{Kasatkin2018} D. V. Kasatkin, and V. I. Nekorkin,
\textit{EPJST} {\bf 227}, 1051–1061 (2018).

\bibitem{Brede2019} M. Brede and A. C. Kalloniatis \textit{Phys. Rev. E}
{\bf 99}, 032303 (2019).

\bibitem{Chen2009} M. Chen, Y. Shang, C. Zhou, Y. Wu, and J. Kurths,
\textit{Chaos} {\bf 19}, 013105 (2009).

\bibitem{Chakravartula2017} S. Chakravartula, P. Indic, B. Sundaram, and
T. Killingback, \textit{Plos One} {\bf 12}, e0178975 (2017).



\bibitem{Niyogi2009} R. K. Niyogi, and L. Q. English, \textit{Phys. Rev.
E} {\bf 80}, 066213 (2009). \bibitem{Skardal2014} P.S. Skardal, D.
Taylor, and J. G. Restrepo, \textit{Physica D} {\bf 267} 27-35 (2014).

\bibitem{Li2010} M. Li, S. Guan, and C-H Lai, \textit{New Journal of
Physics} {\bf 12} 103032 (2010).

\bibitem{Khanra2018} P. Khanra, P. Kundu, C. Hens, and P. Pal,
\textit{Phys. Rev. E} {\bf 98}, 052315 (2018).

\bibitem{Kachhvah2017} A. Kachhvah and S. Jalan, \textit{Euro Phys.
Lett.} {\bf 119}, 60005 (2017).

\bibitem{Danziger2019} M. M. Danziger, I. Bonamassa, S. Boccaletti, and
S. Havlin, \textit{Nat. Phys} {\bf 15} 178–185 (2019).

\bibitem{Danziger2016} M. M. Danziger, O. I. Moskalenko, S. A. Kurkin,
X. Zhang, S. Havlin, and S. Boccaletti, \textit{Chaos} {\bf 26} 065307
(2016).

\bibitem{Pinto2015} R. S. Pinto and A. Saa, \textit{Phys. Rev. E} {\bf
92}, 062801 (2015). \bibitem{Jesus2011} J. G\'{o}mez-Garde\~{n}es, S.
Gomez, A. Arenas, and Y. Moreno, \textit{Phys.Rev. Lett.} {\bf 106},
128701 (2011). P. Ji, T. Peron, P. J. Menck, F. A. Rodrigues, and J.
Kurths, \textit{Phys.Rev. Lett.} {\bf 110}, 218701 (2013)

\bibitem{Boccaletti2016} S. Boccaletti, J.A. Almendral, S. Guan, I.
Leyva, Z. Liu, I. Sendi{\~n}a-Nadal, Z. Wan, and Y. Zou, \textit{ Phys.
Rep.} {\bf 660}, 1-94 (2016);

\bibitem{Leyva2013}I. Leyva, A. Navas, I. Sendi\~{n}a-Nadal, J. A.
Almendral,  J. M. Buldu, M. Zanin, D. Papo, and S. Boccaletti,
\textit{Scientific Reports} {\bf 3} 1281 (2013).  Y. Zou, T. Pereira, M.
Small, Z. Liu, and J. Kurths,  \textit {Phys. Rev. Lett.} {\bf 112}
114102 (2014).

\bibitem{Filatrella2007} G. Filatrella, N. F. Pedersen, and K.
Wiesenfeld, \textit{Phys. Rev. E} {\bf 75}, 017201 (2007).

\bibitem{Zhang2015} X. Zhang, S. Boccaletti, S. Guan, and Z. Liu,
\textit{Phys. Rev. Lett.} {\bf 114} 038701 (2015).

\bibitem{Kuramoto1984} Y. Kuramoto, \textit{Chemical Oscillations,
Waves, and Turbulence}, (Springer, New York, 1984).

\bibitem{Acebron2015} J. A. Acebr{\`o}n, L. L. Bonilla, C. J. P.
Vicente, F. Ritort, and R. Spigler, \textit{ Rev. Mod. Phys.} {\bf 77},
137-185 (2005). \bibitem{Dorfler2012} F. D{\"o}rfler and F. Bullo, 
\textit{SIAM Journal on Control and Optimization} {\bf 50}(3), 1616-1642
(2012).

\bibitem{Rodrigues2016}F. A. Rodrigues, T. K. D. M. Peron, P. Ji, and J.
Kurths, \textit{ Phys. Rep.} {\bf 610}, 1-98 (2016).

\bibitem{Sakaguchi1986} H. Sakaguchi and Y. Kuramoto, \textit{Prog.
Theor. Phys.} {\bf 76}, 576 (1986). \bibitem{Omel'chenko2012}O. E.
Omel'chenko and M. Wolfrum, \textit{ Phys. Rev. Lett.} {\bf 109}, 164101
(2012).

\bibitem{Lohe2015} M. A. Lohe, \textit{Automatica} {\bf 54}, 114-123
(2015).

\bibitem{Skardal2015} P. S. Skardal, D. Taylor, J. Sun, and A. Arenas,
\textit{Phys. Rev. E} {\bf 91}, 010802(R) (2015).

\bibitem{Skardal2015_PhysD} P. S. Skardal, D. Taylor, J. Sun, and A.
Arenas, \textit{Physica D: Nonlinear Phenomena} {\bf 323}, 40-48 (2015).

\bibitem{Ott2008} E. Ott, and T. M. Antonsen, \textit{Chaos} {\bf 18},
037113 (2008). \bibitem{Peron2012} T. K. D. Peron and F. A. Rodrigues,
\textit{Phys. Rev. E} {\bf 86}, 016102 (2012). \bibitem{Ichinomiya2004}
T. Ichinomiya, \textit{Phys. Rev. E} {\bf 70}, 026116 (2004).
\bibitem{Dorogovtsev2008} S. N. Dorogovtsev, A. V. Goltsev, and J. F. F.
Mendes, \textit{Rev. Mod. Phys.} {\bf 80}, 1275 (2008).
\bibitem{Coutinho2013} B. C. Coutinho, A. V. Goltsev, S. N. Dorogovtsev,
and J. F. F. Mendes, \textit{Phys. Rev. E} {\bf 87}, 032106 (2013);  S.
Yoon, M. Sorbaro Sindaci, A. V. Goltsev, and J. F. F. Mendes,
\textit{Phys. Rev. E} {\bf 91}, 032814 (2015).

\bibitem{Kundu2017} P. Kundu, P. Khanra, C. Hens, and P. Pal,
\textit{Phys. Rev. E} {\bf 96}, 052216 (2017). 
%
%
\bibitem{Kundu2018} P. Kundu, C. Hens, B. Barzel, and P. Pal,
\textit{Europhys. Lett.} {\bf 120}, 40002 (2018). \bibitem{Brede2016} M.
Brede and A. C. Kalloniatis, \textit{Phys. Rev. E} {\bf 93}, 062315
(2016).



\bibitem{ana} A. Ben-Israel, T. N. E. Greville, \textit{Generalized
Inverses Theory and Applications}, (Canadian Mathematical Society,
Springer, 2nd edition, 2002).
%
%
%
%
\end{thebibliography}
 \end{document}